\documentclass[letterpaper,twocolumn,fleqn]{article} 

\usepackage{ist}
\usepackage[latin9]{inputenc}
\usepackage{amsmath}
\usepackage{graphicx}
\usepackage{multirow}
\usepackage{soul,color}

\usepackage{nameref}
\usepackage{varioref}
\usepackage{hyperref}
\usepackage{cleveref}

\definecolor{lightblue}{rgb}{.4,1,1}

\definecolor{lightpink}{rgb}{1,.9,.9}

\DeclareRobustCommand*{\IEEEauthorrefmark}[1]{%
  \raisebox{0pt}[0pt][0pt]{\textsuperscript{\footnotesize #1}}%
}

\title{PRNU Estimation from Encoded Videos Using Block-Based Weighting}

\author{{Enes Altinisik\IEEEauthorrefmark{1},
Kasim Tasdemir\IEEEauthorrefmark{2},
H\"usrev Taha Sencar\IEEEauthorrefmark{1}
}
\\
{\IEEEauthorrefmark{1}Qatar Computing Research Institute, HBKU, Qatar; } 
{\IEEEauthorrefmark{2}Abdullah Gul University, Kayseri, Turkey}
}

\date{} 

\begin{document}

\maketitle 

\thispagestyle{empty} 

\begin{abstract}

Estimating the photo-response non-uniformity (PRNU) of an imaging sensor from videos is a challenging task due to complications created by several processing steps in the camera imaging pipeline.
Among these steps, video coding is one of the most disruptive to PRNU estimation because of its lossy nature.
Since videos are always stored in a compressed format, the ability to cope with the disruptive effects of encoding is central to reliable attribution.
In this work, by focusing on the block-based operation of widely used video coding standards, we present an improved approach to PRNU estimation that exploits this behavior.
To this purpose, several PRNU weighting schemes that utilize block-level parameters, such as encoding block type, quantization strength, and rate-distortion value, are proposed and compared.
Our results show that the use of the coding rate of a block serves as a better estimator for the strength of PRNU with almost three times improvement in the matching statistic at low to medium coding bitrates as compared to the basic estimation method developed for photos.
\end{abstract}


%


\section{Introduction}

Videos pose several challenges to estimating the photo-response non-uniformity (PRNU) of imaging sensors as compared to photos. 
Capturing a video not only requires handling a large amount of sensor data but also involves several additional post processing steps for quality improvement. 
In this regard, downsizing, image stabilization, and video compression are three commonly deployed processing blocks in camera pipelines.
In terms of their impact, however, these operations are very detrimental to PRNU estimation and matching as they introduce geometric transformations that distort pixel-to-pixel correspondences in unknown ways and incur information loss that weakens the underlying PRNU pattern.    
Several solutions have been proposed to deal with the adverse effects of these operations; however, PRNU based attribution of videos remains to be a complex task requiring further improvements. 

During video acquisition cameras downsize the high resolution imaging sensor readout to create lower dimensional pictures through a combination of scaling and cropping. 
Since the number of resolutions supported by cameras is quite large \cite{ei2019}, dimension mismatches between the reference PRNU pattern and the video whose source is in question are inevitable. 
Therefore when attributing a video to its source, downsizing parameters need to be determined first. 
Several works have examined in-camera downsizing behavior and introduced procedures to identify the necessary parameters \cite{ei2019, piva, taspinar2019source}. 
One difficulty in applying these approaches is that when combined with the effects of other processing, such as image stabilization, 
correct identification of these parameters become computationally challenging, thereby making PRNU matching more error prone.
Although these parameters can be determined for each camera model in advance from videos captured under constrained settings, 
creating a comprehensive dictionary of downsizing parameters for potentially thousands of camera models requires a significant effort.

Image stabilization is another intermediate processing that video frames undergo to eliminate unwanted camera motion.
When performed electronically, stabilization potentially subjects each frame to a different geometric transformation. 
There are several video stabilization with implementation level specifics typically unknown.
Stabilization creates a significant complication for PRNU based source attribution as the applied transformations need to be determined blindly for each frame separately.
A number of approaches have been proposed considering application of frame-level affine transformations  \cite{piva, taspinar2016,luisapaper} and spatially-variant transformations \cite{altinisik2019}. 
These approaches, in common, perform a brute search of transformation parameters; therefore, they are computationally demanding.
The main difficulty of dealing with stabilized videos, however, remains to be the weakened PRNU pattern in pictures of a video due to overall processing which makes this search very susceptible to errors. 


At the final stage of the camera pipeline, the captured sequence of frames are encoded into a compressed video file.
Video compression is much more sophisticated than still image coding as it is based on motion prediction and defines different types of frames and blocks in its operation.  
As a consequence, the weakening effect of compression on the PRNU pattern in a typical video frame is much higher than that in a typical photo \cite{H264H265}. 
Earlier work in this field mainly focused on combatting effects of a filtering operation, namely, the loop filtering, performed to suppress blockiness artifact introduced by video encoder by trying to eliminate its effects on the PRNU pattern as much as possible \cite{PRNUvideoJessica,MACEfilter} or utilizing only intra-coded frames for PRNU estimation \cite{IPBdiff}.  
Later, it has been shown that the adverse effects of the loop filter can be more effectively compensated at the decoder, rather than in post-processing, \cite{eusipco18}. 

Another characteristic of video coding is that the specifics of quantization operation applied to prediction error of a block is decided at a block level with parameters changing from one block to another to typically achieve a target bitrate. 
To exploit this behavior, recent work also focused on treating the contribution of each block differently.
In \cite{DirikRecent}, authors proposed the elimination of blocks that lack high frequency content in the prediction error from PRNU estimation.
Alternatively,  \cite{H264H265} examined the relation between quantization parameter of a block and the reliability of the PRNU pattern to introduce a block level weighting scheme that adjusts the contribution of each block accordingly. 

In this work, we aim at further improving these approaches by better utilizing block level decoding parameters available at the decoder and consider different weighting schemes to more reliably estimate the PRNU pattern. 
To this purpose, we expand on the earlier approach that utilizes quantization parameter as an estimator for the strength of PRNU pattern by incorporating the rate-distortion cost of coding block data to PRNU estimation.
In addition, we investigate the impact of the so-called skipped blocks, which refer to minimally coded blocks whose parameters are essentially copied from its spatially preceding non-skipped block.
We must note that block-based PRNU weighting schemes disregard the picture content and only utilize decoding parameters.
Hence, the approaches proposed to enhance the strength of PRNU by supressing content interference \cite{li2010source, kang2011enhancing, lin2016enhancing} and to improve the denoising performance \cite{al2015novel, zeng2016fast, kirchner2019spn} can be further incorporated with the introduced weighting schemes. 
Overall, our results show that the coding rate of a block serves as a much better estimator for the quality of the PRNU pattern.

The rest of the paper is organized as follows: First, we examine the H.264 coding standard with a focus on its block level operation. 
Then, block-based PRNU estimation approaches are introduced to mitigate video coding artifacts.
This is followed by a comparison of source camera attribution results and a summary of key findings.

\section{Block-Level Video Coding}
\label{sec:coding}
H.264/AVC {~\cite{kitap}} is one of the most widely used video coding standard today.
At its core, this block-based coding standard leverages the fact that pictures of a video are highly correlated and try to minimize the temporal redundancy between the successive frames.
An encoded video essentially comprises a sequence of frames where  a frame is a conceptual structure that contains information about the compression parameters and the data needed to reconstruct a picture.
The compression of a frame might depend on other frames. In this regard, \textit{P} type frames depend only on a past frame whereas \textit{B} type frames might depend on the past and future frames. In contrast, \textit{I} type frames are compressed independently of others to prevent a possible error propagation in case of a data transmission error.

During encoding the picture of a frame is split into smaller blocks, referred to as \textit{macro blocks}, and each block is compressed individually. 
Just as it is with frames, blocks also have different types. 
A block that depends on a block of a past frame is defined as a \textit{P} block. 
Similarly, a \textit{B} block depends on blocks of the past and future frames, and an \textit{I} block depends only on neighboring blocks of the same frame.
During encoding each picture block is predicted either from the neighbouring blocks of the same frame (intra-prediction) or from the blocks of the past or future frames (inter-prediction). 
Therefore, each coded block contains the position information of the reference picture (motion vector) and the difference between the current block and the reference picture block (residual). This residual matrix is first transformed to the frequency domain and then quantized. 
This is also the stage where information loss takes place.
The severity of the compression and correspondingly the amount of distortion incurred by a block is determined by the choice of the quantization parameter ($QP$) which sets the quantization step size.
Thus, a higher degradation is expected in blocks coded with higher $QP$ values.

Another type of block used during encoding is the {\em skip block}.
A skip block essentially signals the decoder that a block has the same motion vectors and parameters ({\em i.e.}, the quantization parameter and the quantized transform coefficients) as its spatially preceding block. 
In other words, skipped blocks do not contain any pixel information, and it takes only a few bits to encode them.
As a consequence, among all types of blocks, skipped blocks introduce the highest distortion to the reconstructed block.
Therefore, they are of more concern for the PRNU estimation. 


There is ultimately a trade-off between the amount of data used for representing a picture (data rate) and the quality of the reconstructed picture.
Finding an optimum balance has been a research topic for a long time in video coding. The approach taken by H.264  standard is based on rate-distortion optimization (RDO). 
According to the RDO algorithm, the relation between the rate and distortion is defined as  
\begin{equation}
\label{eq:RDO}
    J=D+\lambda R,
\end{equation}
where $D$ is the distortion introduced to a picture block, $R$ is the number of bits required for its coding, $\lambda $ is a Lagrangian multiplier computed as a function of the $QP$ ({\em i.e.}, $\lambda=0.852^{\frac{QP-12}{3}}$ \cite{kitap}), and $J$ is the rate-distortion (RD) value to be minimized. The value of $J$ obviously depends on several coding parameters such as block type (I, P or B), intra-prediction mode (samples used for extrapolation), number of pair of motion vectors, sub-block size (4x4, 8x4, {\em etc.}), and the $QP$. 
Each parameter combination is called \textit{coding mode} of the block, and the encoder picks the mode $m$ that yields the minimum $J$ as given in
\begin{equation}
\label{eq:Jmin}
    m = \underset{i}{\mathrm{argmin}\ J_i}
\end{equation}
where the index $i$ ranges over all possible coding modes.
It must be noted that the selection for the optimum mode in Eq. (\ref{eq:Jmin}) is performed on the encoder side. 
That is, the encoder knows all variables involved in computation of Eq. (\ref{eq:RDO}).
The decoder, in contrast, has only the knowledge of the rate ($R$) and the $\lambda $ value and oblivious to $D$ and $J$ values.

The reliability of an estimated PRNU pattern from a picture of a video evidently depends on the amount of distortion introduced to it during coding. 
Since encoding induced distortion $D$ varies at the block level, its knowledge might be extremely useful during PRNU estimation in identifying the highly distorted blocks.
That is, unlike in photos where the strength of quantization can be used as an estimator for the strength of the PRNU pattern, the $D$ value associated with each block of a picture will serve as a better estimator for the strength of the PRNU pattern than the corresponding $QP$ value.
The distortion information, however, is not available during decoding.
Therefore, in the lack of this information, one can alternatively consider using $R$ to more reliably estimate PRNU.

\section{PRNU Weighting Schemes}
\label{sec:methods}
Given a video compressed by H.264 (or the newer video coding standard H.265~\cite{H265}), the first step of source attribution is the elimination of the loop filtering step at the decoder as described in {\cite{H264H265}}. 
The adverse effects of encoding related information loss must then be countered. 
Since encoding is performed at the block-level, its weakening effect on the PRNU pattern must also be compensated at the block-level by essentially weighting the contribution of each block in accordance with the level of distortion introduced to it.

When the video is not stabilized (or very lightly stabilized) this can be incorporated into conventional PRNU estimation method \cite{chen} by introducing a block-wise mask for each picture as given below
\begin{equation}
K=\dfrac{\sum_{i=1}^{N} I_{i} \times W_{i} \times M_{i}}{\sum_{i=1}^{N} (I_{i})^2 \times M_{i}}
\label{eq2}
\end{equation}
where $K$ is the sensor specific PRNU factor computed using a set of video pictures $I_1,\ldots,I_N$; $W_i$ represents the noise residue obtained after denoising picture $I_i$; and $M_{i}$ is a mask to appropriately weight contribution of each block based on different block-level parameters. 
When the video is stabilized, however, Eq. (\ref{eq2}) cannot be utilized directly since each picture would have been subjected to a different stabilization transformation. 
Therefore, each frame has to be the first inverse transformed and Eq. (\ref{eq2}) must be evaluated after applying the identified inverse transformation to both $I_i$ and the mask $M_{i}$.

When creating a mask $M$, several weighting schemes can be devised depending on which block-level coding parameters are used. 
In this study, we consider four PRNU weighting schemes that essentially depend on masking out skipped blocks, the quantization strength, a hybrid approach that combines both, and the rate-distortion value. 
Developing a formulation that relates such block-level parameters directly to the strength of the PRNU is in general analytically intractable.
Therefore, we need to rely on observations obtained by changing coding parameters in a controlled manner and determine how the estimated PRNU pattern affects the accuracy of the attribution. 
To this end, we use a set of very high quality, almost uncompressed videos and encoded them with different parameters when determining the specifics of each weighting scheme.


These videos are captured using various Android smartphone cameras through a custom built camera app that allows capturing videos at the highest possible bitrate  ({\em i.e.,} corresponding to $QP=1$) while turning off stabilization and electronic zoom to correctly determine the weighting function without interference from other in-camera post-processing \cite{ei2019}. 
All videos include indoor scenes captured  under  natural light by moving the cameras at the highest supported frame resolution of the camera by limiting the duration of each video to 4 seconds.
One video from each camera is used to obtain a reference PRNU pattern of the sensor and another video is used for tests.
In all cases, PRNU patterns are obtained using the loop-filter compensated versions of the pictures extracted from each video.

The test videos are re-encoded using H.264 encoder at different bitrates and quantization parameters depending on the requirements of each scheme as described below. 
When evaluating the goodness of a block-level weighting scheme, we use the conventional peak-to-correlation energy (PCE) detection statistic computed between the reference PRNU pattern and PRNU patterns obtained from pictures of videos using that scheme.

\subsubsection{Eliminating Skipped Blocks}

Skip blocks do not transmit any block residual information of their own. 
Therefore they provide large bitrate savings at the expense of a much higher distortion when compared to other block types as implied by Eq. (\ref{eq:RDO}).
As a consequence, the PRNU pattern extracted from skipped blocks will be the least reliable. 
(It must be noted that the distortion introduced by substituting a skip block with a reference block cannot be arbitrarily high as it is bounded by the RDO algorithm.)
This weighting scheme, therefore, deploys a binary mask $M$ where all pixel locations corresponding to a skipped block are set to zero and those corresponding to coded blocks are assigned a value of one.  
A major concern with the use of this scheme is that at lower bitrates, skip blocks are expected to be much more frequently encountered than other types of blocks, thereby potentially leaving very few blocks to reliably estimate a PRNU pattern.

\begin{figure}[!h]
	\centering
	\includegraphics[width=1\columnwidth]{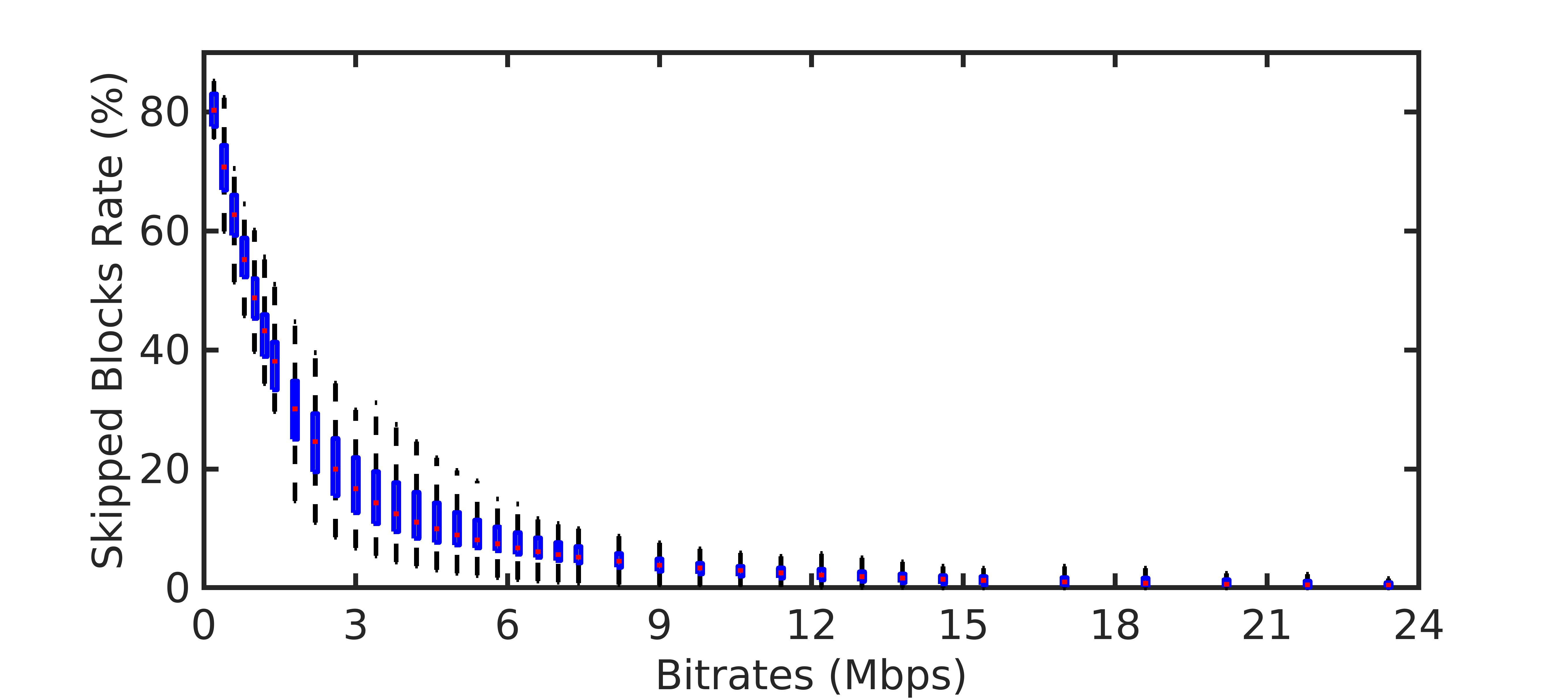}
	\caption{The box plot of SBR measurements from $740$ videos created by re-encoding 20 uncompressed videos at $37$ different bitrates.}
	\label{fig:skipped}
\end{figure}

To examine the behavior at low bitrates, we performed a measurement using 20 uncompressed ($QP=1$) videos captured by different cameras.
These videos are encoded repeatedly at 37 different bitrates from 200 Kbps to 2.3 Mbps. 
This yielded a total of $740$ videos with different amount of skipped blocks depending on the frame resolution, video content, and the bitrate.
Then, for all videos skipped block rate (SBR) is determined as the ratio of the number of skipped blocks to the total number of blocks in each video.
The resulting SBR-bitrate relation is presented in Fig. \ref{fig:skipped}.
Accordingly, it is determined that for videos with a bitrate lower than 1 Mbps, more than 70\% of the blocks are skipped during coding. 
This finding overall indicates that unless the video is extremely short in duration or encoded at a very low bitrate, elimination of skipped blocks will not significantly impair the PRNU estimation.

\subsubsection{Quantization Based PRNU Weighting}

Quantization of the prediction residue associated with a block is the main reason behind the weakening of the PRNU pattern during compression.
Therefore in \cite{H264H265}, the strength of quantization is used as the basis of the weighting scheme.
Since the $QP$ associated with each block is available at the decoder, this work empirically obtained a relation between $QP$ and PCE. 
The underlying relation is obtained by re-encoding 28 high-quality videos at all possible,{\em i.e.,} 51, QP levels. 
Then, average PCE values between video pictures coded at different $QP$ values and the camera reference pattern is computed separately for each video. 
(It must be noted here that setting a fixed $QP$ value for all the blocks in a video does not prevent encoder to use skip blocks.)
To suppress the camera dependent variation in PCE values, obtained average PCE values are normalized with respect to the PCE value obtained for the $QP$ value of 15.
Finally, resulting normalized average PCE values are further averaged across all videos to obtain a camera independent relationship between PCE and $QP$ values.

Translating the obtained $QP$-PCE relation into a weighting function requires taking into account the formulation of the PCE metric. 
Given a reference and estimated PRNU pattern, PCE is defined as the ratio between the square of the correlation between the two patterns and the total energy in the cross-correlation plane except around a spot around the peak correlation.
Essentially, the term in the denominator of this ratio can be interpreted as noise due to pixel-wise independent nature of PRNU, regardless of whether the two estimates match or not. 
Hence, the ratio of two PCE values corresponding to PRNU factors estimated from (two) videos captured the by same source camera can be viewed as the square of the ratio of the energies of the two PRNU patterns \cite{H264H265}.
Overall, taking the square root of the $QP$-PCE curve 
yields the needed weighting functions as shown in Fig. \ref{fig:weight} (red colored curve). 
Correspondingly, the mask $M$ in Eq. (\ref{eq2}) is obtained by filling all pixel locations corresponding to a block with a particular $QP$ value with the respective values using this curve.

\subsubsection {Quantization Based PRNU Weighting Without Skipped Blocks} 
Since skipped blocks introduce higher distortion, they can be eliminated from PRNU estimation as discussed above. 
However, the contribution of remaining, coded blocks can still be weighted in accordance with their distortion.
Hence, another weighting scheme that follows the previous two schemes is the quantization based weighting of non-skipped blocks. 
This can be simply realized by limiting the above described procedure devised considering all blocks to only coded blocks and obtaining the $QP$-PCE relationship and the corresponding weighting function. 

\begin{figure}[!h]
	\centering
	\includegraphics[width=0.9\columnwidth]{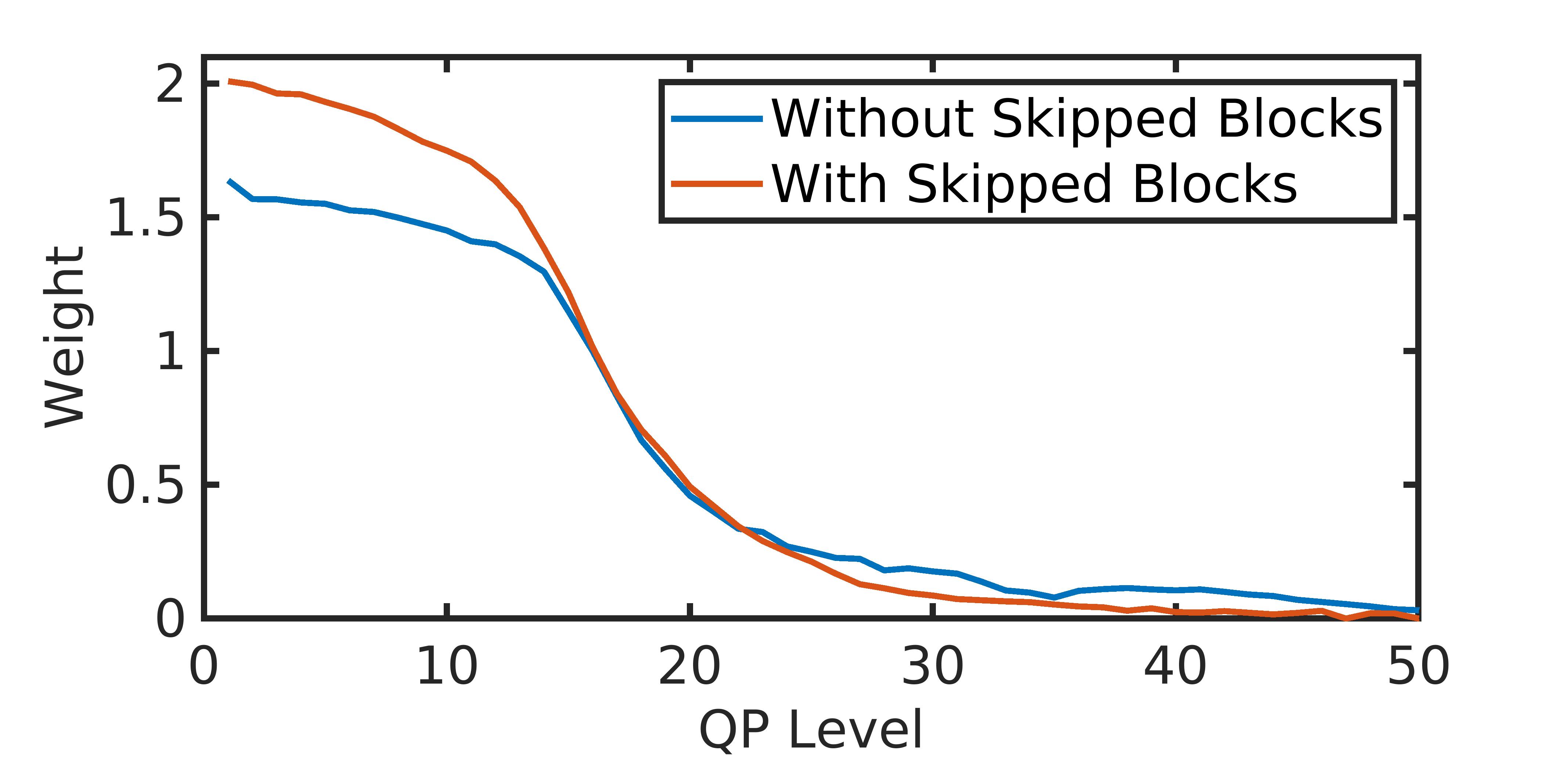}
	\caption{Quantization based weighting functions for PRNU blocks depending on $QP$ values when skip blocks are included (red curve) and eliminated (blue curve).}
	\label{fig:weight}
\end{figure}

Repeating the same process using videos captured by 20 cameras while eliminating all skipped blocks, the weighting function given in Fig. \ref{fig:weight} (blue colored curve) is obtained.
As can be seen, both curves exhibit similar characteristics; however, the latter one exhibits a lesser change in weights depending on $QP$, which indicates that the variation in the strength of the PRNU estimated across all coded blocks is less variable. 
This is mainly because eliminating skipped blocks yields higher PCE values, and this increase is more apparent at high $QP$ values where compression is more severe and skip blocks are more frequent.
Therefore, when the resulting $QP$-PCE values are normalized with respect to a fixed PCE value ($QP=15$), it yields a more compact weight function that takes relatively lower weights at low $QP$ levels and higher weights at high $QP$ levels.
This new weight function is used in a similar way when creating a mask for each video picture with one difference that for skipped blocks the corresponding mask values are set to zero regardless of the $QP$ value of the block.


\subsubsection{RDO Based PRNU Weighting}




Although one can infer the distortion introduced to a block in terms of its type and/or through the strength of quantization ($QP$), the correct evaluation of the reliability of the PRNU ultimately depends on the actual distortion. 
Therefore incorporating RDO algorithm into PRNU estimation offers an improved ability to determine PRNU noise quality; however, the variables $D$ and $J$, in Eq. (\ref{eq:RDO}), are unknown at the decoder.
One way to cope with this problem is by assuming that the value of the optimum $J$ for each block does not vary much throughout the video and 
obtaining the relation between the distortion and the rate experimentally.
This can be simply performed by computing the mean square error (MSE) between the original and decoded versions of a block and determining the number of bits spent for coding that block.
In this way $D$ can be estimated from the known $\lambda R$ and than this estimated $D$ value can be used to determine the associated weight for each block.  
However, since the assumption about stability of $J$ is not well-founded, the estimation of the distortion $D$ associated with a block from $QP$ and $R$ values will inevitably introduce some errors.
%
%
%
Therefore, as an alternative, we investigate the feasibility of a PRNU weighting scheme based solely on $\lambda R$, which is known at the decoder just like $QP$.
This essentially requires determining a weighting function that defines the relation between $\lambda R$ and $PCE$ which can be subsequently used for creating a mask for weighting PRNU blocks. 
This relation can be determined empirically through a number of tests as it was done for the quantization based weighting scheme. .

\begin{figure}[htbp]
    \centering
    \includegraphics[width=0.9\columnwidth]{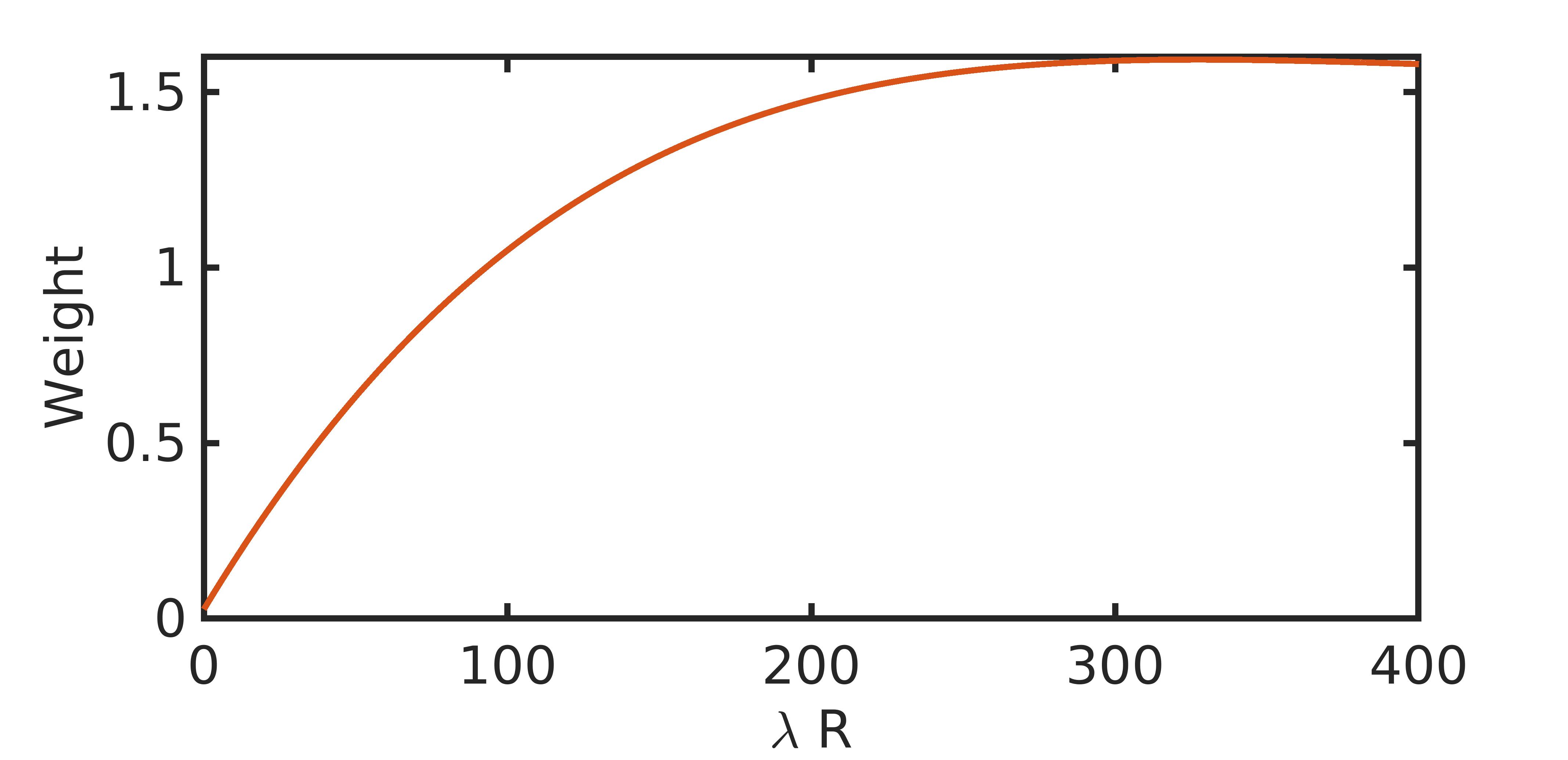}
  \caption{The weighting function based on $\lambda \times R$ using pictures of 20 re-encoded videos at 42 different bitrates.}
    \label{fig:weights}
\end{figure}

With this objective, 20 high-quality videos captured by 20 cameras are re-encoded at 42 different bitrates that vary between 200 Kbps and 32 Mbps.
As for obtaining $\lambda R$-PCE relationship, $\lambda R$ values are measured at a block-level. 
However, the same approach cannot be taken for PCE values because computing PCE at a block-level is not meaningful.
This ideally requires pictures comprising blocks that are encoded at the same rate. 
Since this is not possible, we spliced together new frames in the PRNU domain from PRNU blocks of original pictures of each video by sorting them with respect to their $\lambda R$ values while preserving the location of each block in a frame to not impair the matching process. 
That is, blocks with similar $\lambda R$ values in different pictures of a video are transferred to the same location in the newly generated spliced frames.
Then, for all newly created frames in the PRNU domain 
corresponding PCE values are calculated by matching them with the camera's reference pattern. 
To suppress between-cameras variations in the $\lambda R$-PCE relation, obtained curves for each camera are normalized with respect to the PCE value at a fixed $\lambda R$ ($\lambda R=60$) value and all curves are averaged together to obtain a camera-independent $\lambda R$-PCE relation.
Finally, the weighting function that depends on rate, $\lambda R$, is obtained by taking its square root (see page {\pageref{sec:methods}}) as displayed in Fig. {\ref{fig:weights}}.

\section{Performance Comparison}
\label{sec:results}

To determine how different block based PRNU weighting schemes compare against each other, we used another set of 47 videos\footnote{This set of videos can be obtained at https://github.com/VideoPRNUExtractor/Block-BasedWeighter.} captured similarly by the same 20 cameras using the custom camera app \cite{ei2019}.
It must be noted that these videos were not used in the creation of a reference PRNU pattern or during the computation of the weighting functions.
All videos are captured indoors under normal camera motion and had a duration of 4 seconds.
The videos are captured at their native resolutions that varied between $768\times432$ to $2160\times2160$ pixels at the highest quality (by setting $QP=1$). 
These were then similarly re-encoded using H.264 codec at 11 different bitrates, starting from 600 Kbps up to 4 Mbps, creating a total of 517 test videos.

\begin{figure}[!ht]
\centering
  \begin{minipage}[b]{0.99\columnwidth}
    \centering
    \includegraphics[width=1\columnwidth]{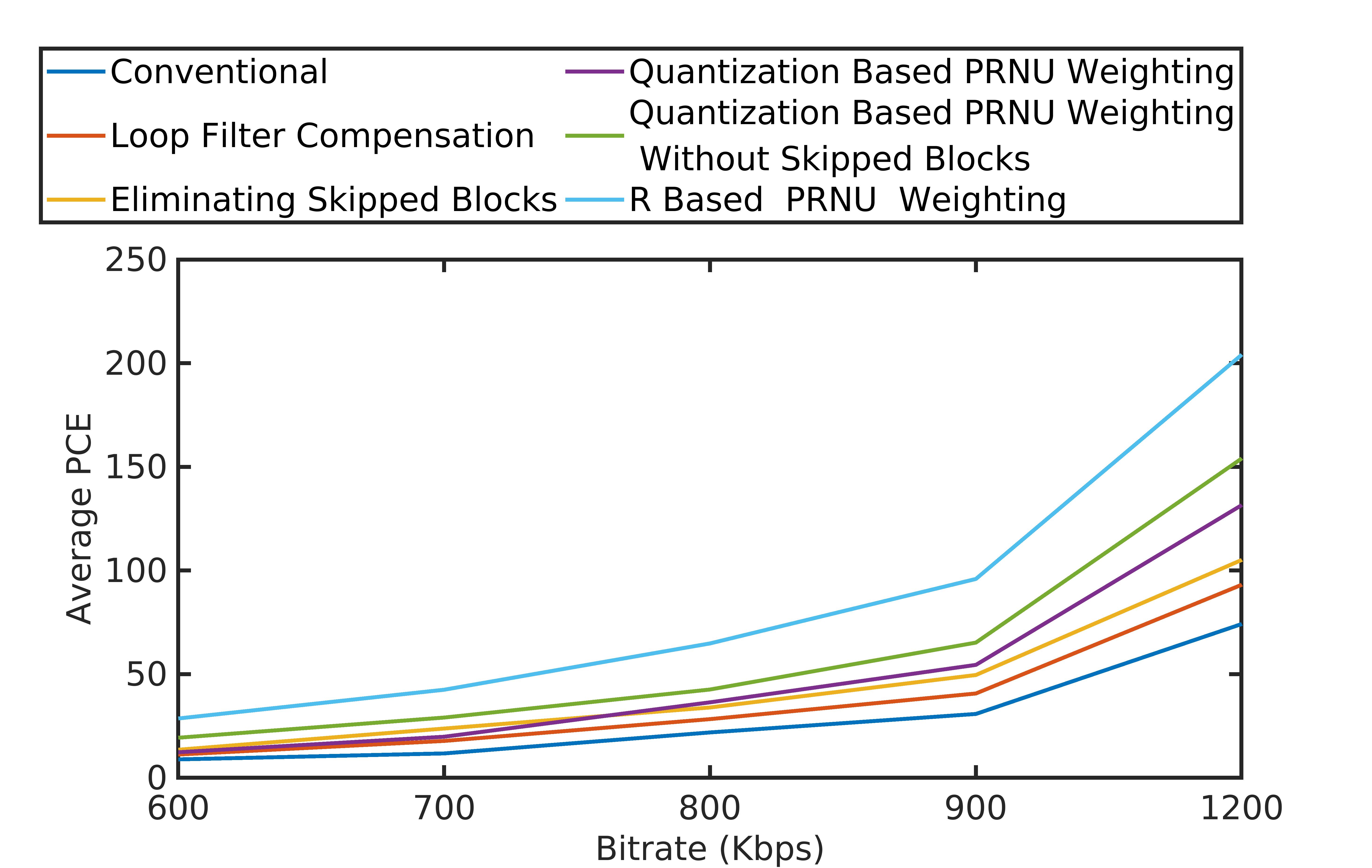}
    \centerline{(a)}
  \end{minipage}
  \centering
  \begin{minipage}[b]{0.99\columnwidth}
    \centering
    \includegraphics[width=1\columnwidth]{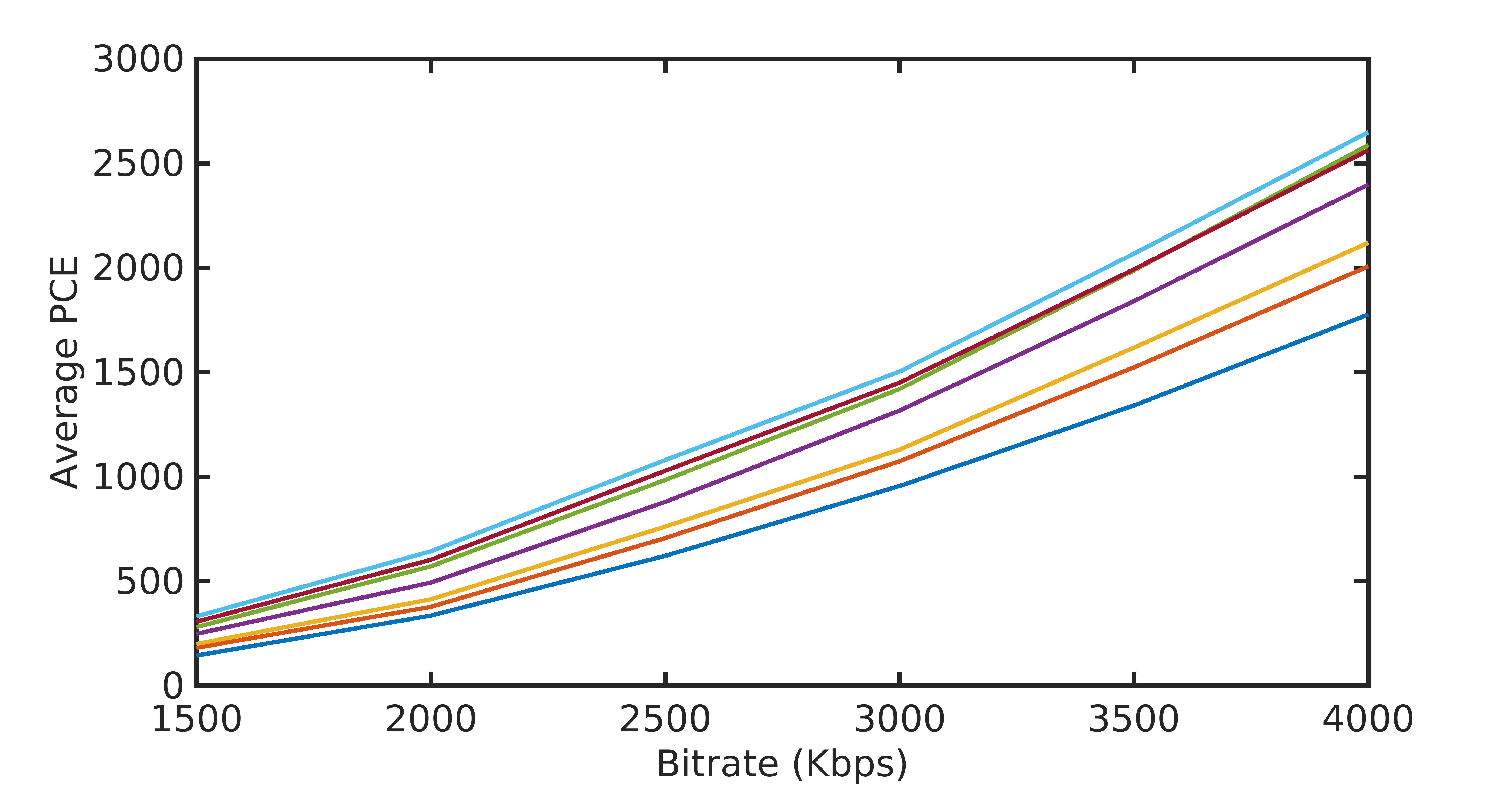}
    \centerline{(b)}
  \end{minipage}
  \caption{Average PCE values for 47 high-quality videos encoded at 11 different bitrates ranging from (a) 600 kbps to 1200 kbps and (b) 1500 kbps to 4000 kbps.}
   \label{fig:result}

 \end{figure}

The PRNU pattern estimated from each video using one of the four weighting schemes, as well as the basic method that treats each video picture and its loop-filter compensated version as a photo, are matched to the reference PRNU pattern of the camera and the PCE values are computed.
Figure \ref{fig:result} presents the average PCE values obtained at low and high bitrates separately for better visibility.
As can be seen from these results, all weighting schemes yield a better estimate of the PRNU as compared to the conventional method or to just compensating the loop filtering. 
In line with our expectations, at all bitrates, rate based PRNU weighting scheme yielded the best results.

It is also observed that at all bitrates the performance of $QP$ based weighting scheme improves when skipped blocks are eliminated from the estimation process.
Although at high bitrates, performances for both of the schemes are expected to converge, there is still a gap at the bitrate of 4 Mbps.
This can be explained by the fact that even at this bitrate 15-20\% of all blocks are skipped as can be seen in Fig. \ref{fig:skipped}. 
It can further be noticed that at 600 and 700 Kbps bitrates elimination of skipped blocks yields slightly better PCE values than quantization based weighting scheme.
This is essentially due to incorrect weighting of skipped blocks, which constitute almost 80\% of all blocks at those bitrates, as their $QP$ values are taken to be those of their reference blocks which are likely to be low.


Results show that at low bitrates (600 to 1200 Kbps) as compared to the conventional method, average PCE values increased by $1.33$ times when only loop filter is compensated; by $1.62$ times if skipped blocks are eliminated; $1.65$ times for $QP$ based PRNU weighting; $2.16$ times for $QP$ based weighting without skipped blocks; and $3.13$ times for $R$ based weighting.
At higher bitrates, average PCE values are observed to increase by a factor of $1.15$, $1.24$, $1.45$, $1.60$, and $1.76$, respectively, for the same methods.

In addition to average improvement in the PCE values, we also determined the number of videos that yield a PCE value above 60, which is the commonly used matching threshold for photos.
The table given below divides videos into 5 groups based on the level compression applied to each video, measured in terms of the average number of bits-per-pixel used during encoding.
Similar to results of Fig. {\ref{fig:result},} it can be seen that at low bits-per-pixel values, PRNU weighting approach yields significant improvement in attribution performance over the basic method with rate based weighting scheme performing significantly better than all other schemes. 
One interesting finding here is that at lower than 0.052 bits-per-pixel compression, elimination of skipped blocks performs better than 
QP based weighting schemes.
We believe the main reason for this phenomenon is that QP based PRNU weighting function could not be reliably determined at high QP values,
as indicated by fluctuations in Fig. \ref{fig:weight} when $QP>35$.

\renewcommand{\tabcolsep}{3pt}
\begin{table}[!ht]
	\centering
	\caption{Number of Videos that Yielded a PCE Value Higher than 60 for All Weighting Schemes}
	\label{ResultTable}
	\vspace{0.5cm}
	\resizebox{\linewidth}{!}{
	\begin{tabular}{c|c|c|c|c|c|c|c}
		Bits Per & Convent. & Lp. Filter & Eliminate & QP & QP w/out  & $\lambda R$ &\multirow{2}{*}{\bf Total}\\	
		Pixel& Method & Comp. & Skip Bl.& Based & Skip Bl. &Based& \\ \hline
		
		\hline
    $<0.024$&2 &3&10&4&7&10&103\\   \hline
    $<0.052$&27&34&42&32&40&55&104\\   \hline
    $<0.084$&40&47&51&46&53&63&103\\   \hline
    $<0.172$&65&69&72&74&77&82&103\\   \hline
    $>0.172$&92&94&93&97&97&97&104\\   \hline
    \bf Total &226&247&268&253&274&307&517\\   \hline

	\end{tabular}
	}
\end{table}

\begin{figure}[!h]
	\centering
	\includegraphics[width=0.9\columnwidth]{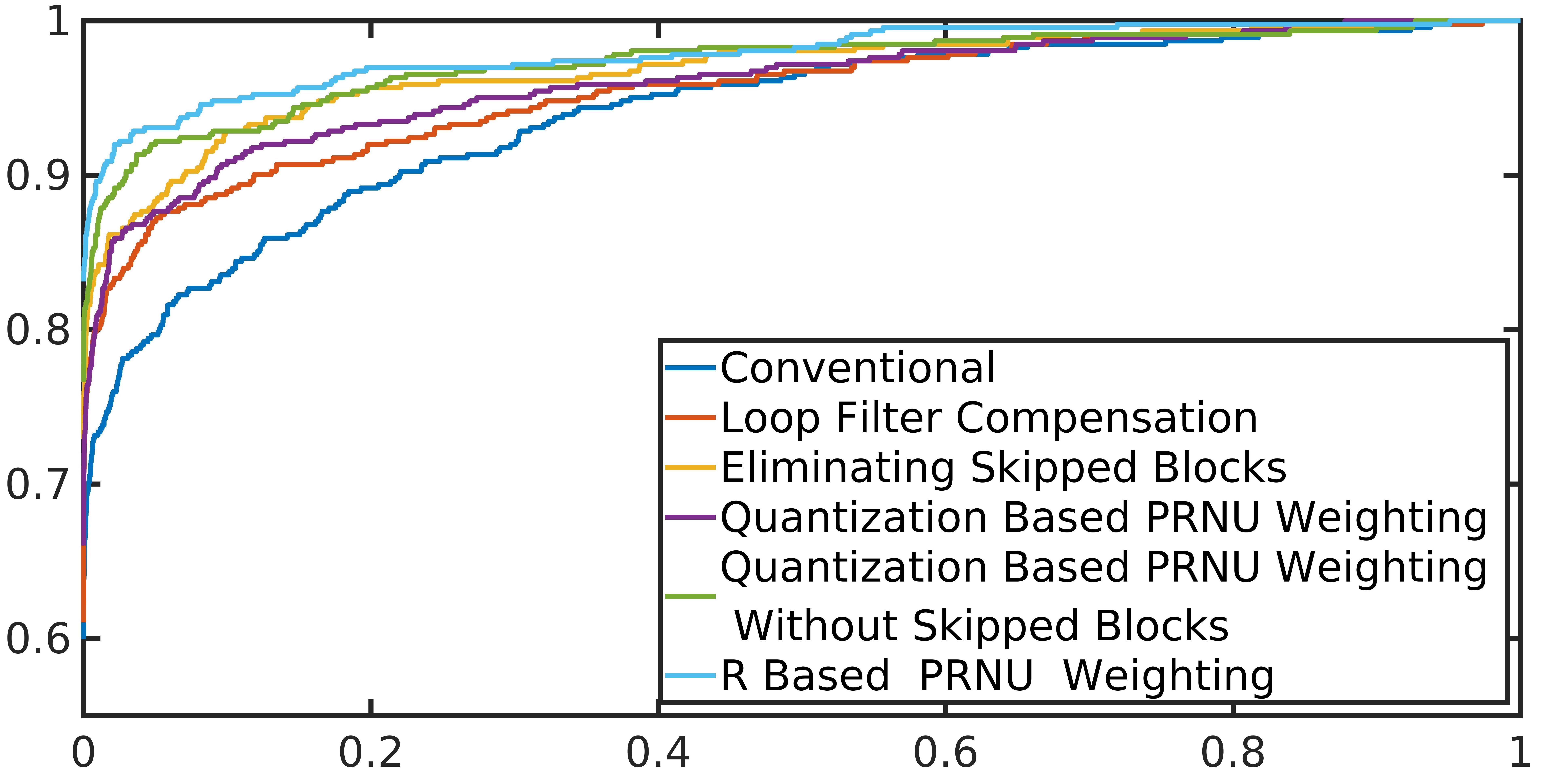}
	\caption{ROC curves corresponding to different PRNU estimation approaches.}
	\label{fig:Roc}
\end{figure}

To ensure that the improvement in the performance due to PRNU weighting does not come at the expense of an increased number of false-positive matches, we performed another experiment by attempting to match each test video to 14 randomly selected non-matching reference patterns. 
This in total yielded $7,238$ PCE values for each weighting scheme.
The PCE values corresponding matching and non-matching reference PRNU patterns are then used to generate receiver operating characteristic (ROC) curves for each of the PRNU weighting schemes as displayed in Fig. \ref{fig:Roc}.

\vspace*{-5mm}

\section{Conclusion}
\label{sec:conclusion}
Overall our results show that when estimating a PRNU pattern from a video, block-level PRNU weighting yields much better results and that the rate associated with a block is a more reliable estimator for the strength of extracted PRNU pattern. 
It is also observed that the improvement in PCE values due to PRNU weighting becomes more visible at bitrates higher than 900 Kbps. 
Our findings, however, also demonstrate that it is a challenge to estimate PRNU pattern from low bitrate videos (600-800 Kbps) even in the absence of other in-camera processing that may interfere with the estimation process. 
Considering a typical video that includes artifacts of other video processing operations, such as downsizing and stabilization, in addition to those of compression, the resulting PCE values will be even lower. 
For those cases, the improvements provided by the PRNU weighting schemes will be more critical for reliable source attribution.

\bibliographystyle{IEEEtran}
\bibliography{IEEEexample}

\end{document}